\documentclass[prl,twocolumn,superscriptaddress,showpacs,longbibliography,a4paper]{revtex4-1}
\pdfoutput=1
\usepackage{amssymb, amsbsy, amsmath, latexsym, dsfont, array, layout, graphicx, mathrsfs, color}

\newcommand{\one}{\mathds{1}}
\newcommand{\ket}[1]{\left|{#1}\right\rangle}
\newcommand{\bra}[1]{\left\langle{#1}\right|}

\newcommand{\ketbrad}[1]{\left|{#1}\rangle\!\langle{#1}\right|}
\newcommand{\ketbra}[2]{\left|{#1}\rangle\!\langle{#2}\right|}

\usepackage[normalem]{ulem}

\usepackage{bm}

\begin{document}

\title{Entanglement-enhanced quantum metrology in a noisy environment}
\author{Kunkun Wang}
\author{Xiaoping Wang}
\author{Xiang Zhan}
\author{Zhihao Bian}
\affiliation{Department of Physics, Southeast University, Nanjing 211189, China}
\affiliation{Beijing Computational Science Research Center, Beijing 100084, China}
\author{Jian Li}
\affiliation{Department of Physics, Southeast University, Nanjing 211189, China}
\author{Barry C. Sanders}
\affiliation{Synergetic Innovation Center in Quantum Information and Quantum Physics, University of Science and Technology of China, CAS, Hefei 230026, China}
\affiliation{Hefei National Laboratory for Physical Sciences at Microscale, University of Science and Technology of China, CAS, Hefei 230026, China}
\affiliation{Institute for Quantum Science and Technology, University of Calgary, Alberta T2N 1N4, Canada}
\affiliation{Program in Quantum Information Science, Canadian Institute for Advanced Research, Toronto, Ontario M5G 1M1, Canada}
\author{Peng Xue}
\email{gnep.eux@gmail.com}
\affiliation{Department of Physics, Southeast University, Nanjing 211189, China}
\affiliation{Beijing Computational Science Research Center, Beijing 100084, China}
\affiliation{State Key Laboratory of Precision Spectroscopy, East China Normal University, Shanghai 200062, China}

\begin{abstract}
Quantum metrology overcomes standard precision limits and plays a central role in science and technology. Practically it is vulnerable to imperfections such as decoherence. Here, we demonstrate quantum metrology for noisy channels such that entanglement with ancillary qubits enhances the quantum Fisher information for phase estimation but not otherwise. Our photonic experiment covers a range of noise for various types of channels, including for two randomly alternating channels such that assisted entanglement fails for each noisy channel individually. We have simulated noisy channels by implementing space-multiplexed dual interferometers with quantum photonic inputs. We have demonstrated the advantage of entanglement-assisted protocols in phase estimation experiment run with either single-probe or multi-probe approach. These results establish that entanglement with ancill\ae\ is a valuable approach for delivering quantum-enhanced metrology. Our new approach to entanglement-assisted quantum metrology via a simple linear-optical interferometric network with easy-to-prepare photonic inputs provides a path towards practical quantum metrology.
\end{abstract}


\maketitle

\noindent

{\it Introduction:-}Quantum metrology ~\cite{VSL04,WGA07,VSL11,RJM12,SMA14,PL17,DAB+17} exploits nonclassicality
to surpass classical limits to interferometric parameter estimation~\cite{SS08,BEJ13,MRC16}.
Quantum metrological enhancement is achieved by employing quantum probes
for detecting physical properties with resolution beyond the reach of classical approaches~\cite{LCF16,LMM17,EIA14,WMF14}.
Without noise,
entangling the measurement system with ancillary quantum degrees of freedom provides no advantage to scaling of measurement precision with number of particles~\cite{AJJ00,VSL06}.
Contrariwise,
in the presence of noise,
which deleteriously affects measurement precision,
entangling with ancill\ae\ is suggested to deliver higher precision than not using entanglement with ancill\ae~\cite{RL14,HMM16,HC17,YF17}.

\begin{figure}
\includegraphics[width=0.4\textwidth]{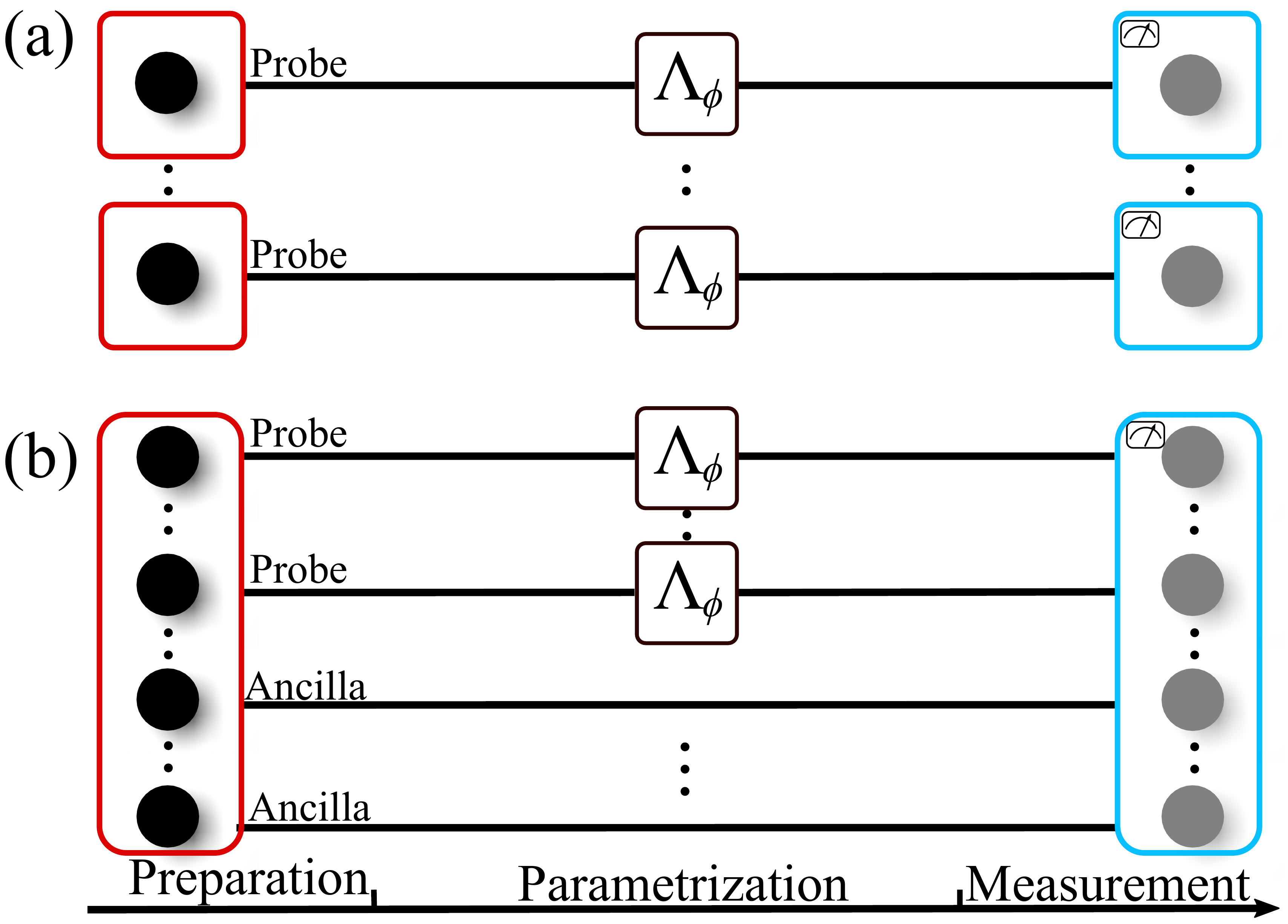}
\caption{Concept of the comparison between the parallel scheme of quantum metrology with and without assisted entanglement. (a) Parallel scheme. Probes go through maps $\Lambda_\phi$ in parallel. (b) Parallel scheme with assisted entanglement. Introducing noiseless ancill\ae\ sharing entanglement with probes, and implementing joint measurements after the evolution give estimation with an enhanced precision.}
\label{circuit}
\end{figure}

We demonstrate experimentally that entangling probes with ancill\ae\ significantly enhances the performance of noisy quantum metrology as quantified by the quantum Fisher information (QFI) for parameter estimation (Fig.~\ref{circuit}). Through entanglement with ancill\ae\, the probe state is less sensitive to noise. Information from probes is limited by the Holevo bound~\cite{H82} whereas enlarging the Hilbert space by entangling with ancill\ae\ allows more information to be accessed by measurements that exploit the larger dimension of Hilbert space. The QFI is obtained by tracing over the auxiliary space, which maximizes over all mixed states. That might make the QFI larger than that without ancill\ae\ \cite{suppl}. The enlargement enhances the precision only for certain noisy channels, for which the input states entangled between the space of probes and ancill\ae\ are optimal~\cite{RJM12,JR13,AH08,F17}.

Based on these theoretical proposals,
we experimentally investigate whether
entangled ancill\ae\ can deliver enhanced metrological precision in the presence of noise~\cite{RJM13,BRL11}
realized as simulated decohering quantum channels~\cite{KRR12,YJO12,Barry17},
and herein establish that indeed entangling with ancill\ae\
is advantageous for efficiently inferring the unknown parameter measuring for a wide range of noise values.
We develop space-multiplexed noisy channels via a dual interferometric network~\cite{KRR12}
and inject hyperentangled photonic states
entangled in their polarizations and spatial modes~\cite{JTP10,ELL10}.

\begin{figure*}
\includegraphics[width=0.9\textwidth]{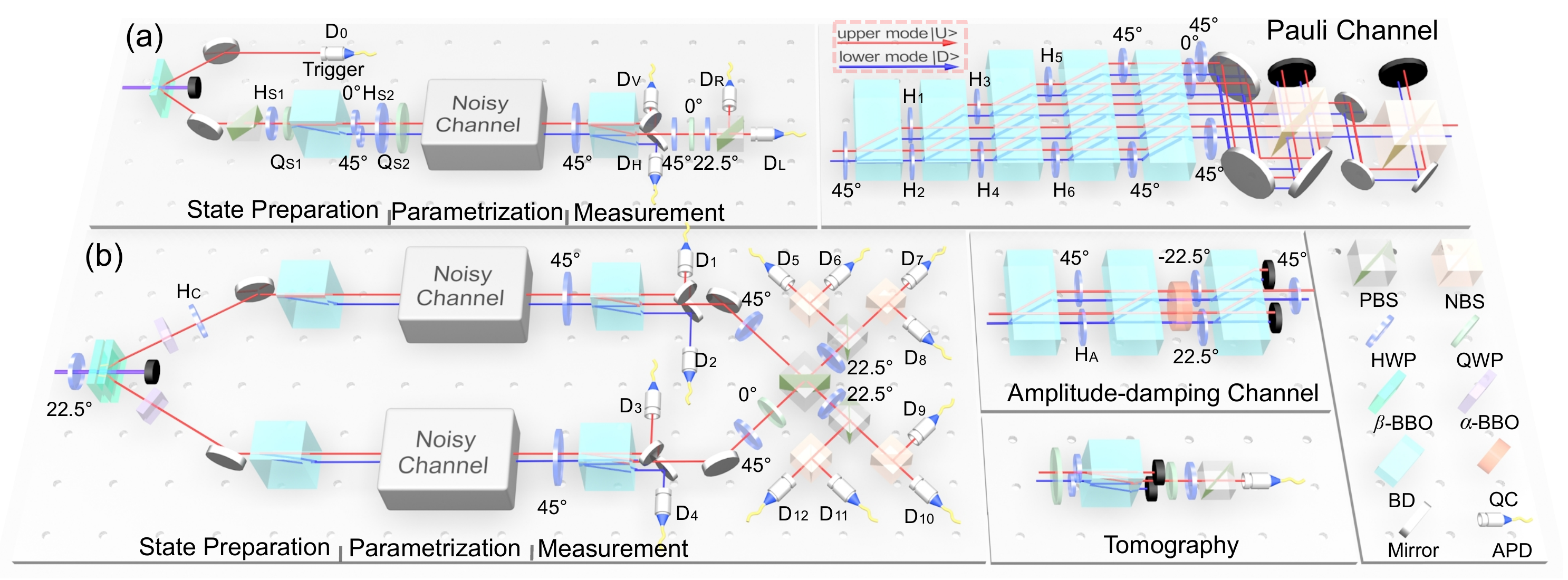}
\caption{Experimental scheme.
(a) Setup for entanglement-assisted single-probe approach. Heralded single photons are used to prepare polarization-spatial hyperentangled states for entanglement-assisted quantum metrology approach. Space-multiplexed noisy channels are realized by the dual interferometric network setup, in which spatial coherence is reduced, and the optical path delay enables the arrival time of the photons passing through different optical paths on the BD (for the amplitude-damping channel) or NBS (for the depolarizing channel) to be different. Random phases are added between photons in different optical paths before recombining them on the BD or NBS. Quantum process tomography is performed via wave plates (WPs), BD and PBS, and enables reconstruction of the process matrices for the channels. (b) Setup for entanglement-assisted two-probe approach. Polarization-entangled photon pairs are used to prepare the four-qubit hyperentangled state. Projective measurements are realized via BDs, WPs, NBSs and a PBS. Coincidences between paired photons are detected by APDs.}
\label{setup}
\end{figure*}

{\it Theory:-}First, we use a single-probe scheme as an example. Entanglement-assisted parameter estimation comprises three stages:
\emph{preparation} in which a probe (a photonic qubit in our case)
shares entanglement with an ancilla;
\emph{parametrization} where the probe evolves in a channel and the parameter to be estimated is encoded in the probe whereas the ancilla does not participate;
and \emph{measurement} in which a joint measurement is performed on both the probe and ancilla to yield
a precise estimate of the parameter.
We focus on a two-level probe
detecting a phase shift modelled by the unitary map
\begin{equation}
	\mathcal{U}_\phi(\rho)
		=U_\phi\rho U_\phi^\dagger,\;
	U_\phi=\ket{0}\bra{0}+\text{e}^{\text{i}\phi}\ket{1}\bra{1}
\end{equation}
for $\rho$ the initial state. The noise map $\mathcal{E}$ acts after $\mathcal{U}_\phi$:
$\phi$ is encoded into the probe state
$\rho_\phi=\Lambda_\phi\rho$
for $\Lambda(\phi)=\mathcal{E}\circ\mathcal{U}_\phi$.

We use QFI~\cite{JMD16}
\begin{equation}
	J(\rho(\phi))=\text{Tr}\left(\rho(\phi) A^2\right),\;
	\frac{\partial\rho(\phi)}{\partial\phi}=\frac{A\rho(\phi)+\rho(\phi) A}{2},
\label{eq:QFI2}
\end{equation}
to quantify the metrological precision,
with~$A$ the symmetric logarithmic-derivative operator.
QFI is an appropriate measure as it serves as an asymptotic measure of the amount of information inherent in how much the system parameters can be acquired by measurement.
The quantum Cram\'{e}r-Rao bound~\cite{SC94} is a lower bound for the precision~$\Delta \phi$ of the estimate of~$\phi$:
$\Delta \phi\geq1/\sqrt{\nu J(\rho(\phi))}$
for~$\nu$ the number of repetitions of the phase-estimate procedure.
The best bound is found by maximizing the QFI,
which depends on both~$\rho$ and~$\phi$.

For a single-probe instance,
noise diminishes the measurement precision
evident through reducing the output-state QFI after passing through~$\mathcal{E}$.
Entangling with an ancilla enhances precision for noisy channels
and the state transformation is $(\Lambda_\phi\otimes\one) \widetilde{\rho}$ with the ancilla unchanged. Here, $\widetilde{\rho}$ denotes the probe+ancilla state whereas~$\rho$ denotes the single-probe state. 

We consider three decoherence processes encountered in quantum-enhanced metrology:
amplitude-damping (spontaneous emission and photon scattering inside the interferometer),
general-Pauli (most general lossless channel)
and depolarizing (most symmetric Pauli channel assuming uncorrelated noise)
channels~\cite{JR13},
which are typically utilized when accounting for decoherence in optical interferometry~\cite{Barry17}.

We start with the amplitude-damping channel~\cite{HMM16}
\begin{equation}
	\sum_{\imath=0}^1
		A_\imath\rho A^\dagger_\imath,\
		A_0=\begin{pmatrix} 1&0\\0&\sqrt{1-\eta}\end{pmatrix},\
		A_1=\begin{pmatrix} 0&\sqrt{\eta}\\0&0\end{pmatrix}
\label{eq:ADC}
\end{equation}
for $\eta$ the probability of decay $\ket{1}\mapsto\ket{0}$.
For a single-probe input state, the optimized QFI is $1-\eta$ and the optimal state is
$\ket+:=(\ket{0}+\ket{1})/\sqrt{2}$. For the entanglement-assisted approach, the QFI is $2(1-\eta)/(2-\eta)$ for an entangled state of the probe and ancilla $\ket{\Phi}:=(\ket{00}+\ket{11})/\sqrt{2}$ and is always greater than that of the case without assisted entanglement for arbitrary $\eta\in(0,1)$~\cite{HMM16}.

For $\bm{\Xi}=(\one,X,Y,Z)$ the Pauli matrices,
the general-Pauli channel is the map
\begin{equation}
	\mathcal{E}_\text{GPC}(\rho)=\sum_{i=0}^3p_i\Xi_i\rho\Xi_i,\
	\sum_i p_i=1,\
	0\leq p_i\leq1,
\label{eq:PC}
\end{equation}
and the depolarizing channel $p_1=p_2=p_3=p/4$ is a special case.
For a single-qubit probe, $\ket+$ is the optimal state,
and the optimal QFI is $(1-p)^2$~\cite{HMM16}.
If the joint-probe ancilla state is $\ket{\Phi}$,
the QFI is $2(1-p)^2/(2-p)$. For arbitrary $p\in(0,1)$, the QFI is always greater than that of the case without assisted entanglement~\cite{HMM16}.

The depolarizing channel can be regarded as a time-sharing combination of a noiseless channel and a noisy channel in which the state will evolve to a maximally mixed state~\cite{RMC+04,SE11,CRV+11}.
For either of the two channels, the entanglement-assisted approach does not provide any advantage.
However, somewhat surprisingly,
assisted entanglement improves QFI for the depolarizing channel.
We can test for the general-Pauli channel (the depolarizing channel is a special case) which can be implemented in a time-sharing way~\cite{RMC+04,SE11,CRV+11,OSP+13,OCM+15}.
Each Pauli operator is applied over a specific activation time, respectively, and the total decoherence process lasted over an activation cycle, achieving a time-sharing general-Pauli channel. To explain the advantages of entanglement-assisted quantum metrology, we rather implement a new type of general-Pauli channel, namely a space-multiplexed Pauli channel.

\begin{figure*}
\includegraphics[width=0.9\textwidth]{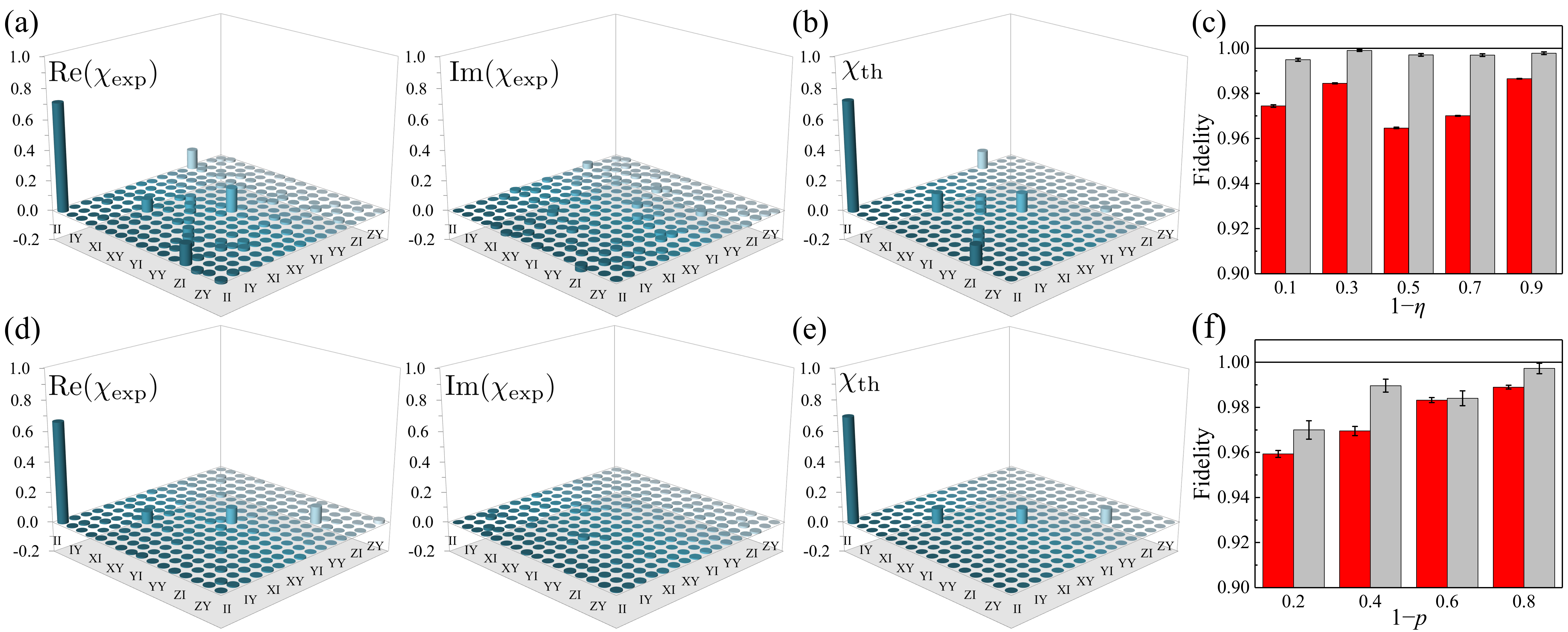}
\caption{Experimental results of the reconstructed noisy channels. For the entanglement-assisted approach, the reconstructed process matrices for the amplitude-damping channel with $\eta=0.5$ (a) and the depolarizing channel with $p=0.4$ (d) compared with their theoretical predictions (b) and (e). The fidelities $F$ of the reconstructed process matrices for the amplitude-damping and depolarizing channels as a function of the noise parameters are shown in (c) and (f), respectively. The red bars indicate the fidelities for the entanglement-assisted approach and the grey ones indicate those for the optimized single-probe approach. Error bars indicate the statistical uncertainty, obtained from Monte-Carlo simulations assuming Poissonian photon-counting statistics.}
\label{channel}
\end{figure*}

Our method can be extended to a more complicated case --- an $N$-probe approach. In the absence of noise, an $N$-probe approach with an optimal $N$-qubit input state (e.g., a N00N state) achieves the Heisenberg limit scaling, which provides improvement over classical limits. However, the advantages are destroyed by noise. Our entanglement-assisted approach in which $N$ probes are entangled with noiseless ancill\ae\ protects against noise and the effect caused by noise can be eliminated by assisted entanglement. Even in the presence of noise, the entanglement-assisted approach beats the shot-noise limit and even maintains the Heisenberg limit scaling for some special noisy channel.

We use a two-probe approach as an example. A two-qubit N00N state (one of the Bell states) $\ket{\Phi^+}=(\ket{00}+\ket{11})/\sqrt{2}$ with both qubits being probes is optimal only in the noiseless case. The phase $\phi$ to be estimated is obtained via the unitary map applied in parallel
\begin{equation}
\mathcal{U}^2_\phi\left(\varrho\right)=U_\phi\otimes U_\phi\varrho U_\phi^\dagger\otimes U_\phi^\dagger
\end{equation}
with $\varrho=\ket{\Phi^+}\bra{\Phi^+}$.
Through a collective noisy channel in parallel, the probe state becomes $\varrho_\phi=\Lambda_\phi^{\otimes2}\varrho$.

A four-qubit entangled state $\tilde{\varrho}=(\ket{0000}+\ket{1111})(\bra{0000}+\bra{1111})/2$ of two probes and two ancill\ae\ beats the optimal state of two probes $\varrho$ in the presence of noise. Taking the collective damping channel as an example, its QFI is
\begin{equation}
\frac{8(\eta-1)^2\{2(\eta-1)^2\cos8\phi+(\eta-2)\eta\left[(\eta-2)\eta+2\right]+2\}}{\left[(\eta-2)\eta+2\right]^3}
\end{equation}
and is larger than that of $\varrho$, even though this particular four-qubit entangled state is not necessarily optimal.

{\it Realization of noisy channels:-}The experimental setup in Fig.~\ref{setup} involves the three stages
of state preparation, parametrization and measurement.
In the preparation stage, we prepare single photons in polarization-spatial hyperentangled states for entanglement-assisted single-probe approach~\cite{JTP10,ELL10}. Whereas, for entanglement-assisted two-probe approach, polarization-entangled photon pairs are used to prepare the four-qubit hyperentangled state~\cite{suppl}.

The probe state is transformed according to the noisy channel, whereas the ancilla qubit is not evolving. The noise is introduced in a controlled way only on the probe. The efficiency of the optimal estimation is shown to outperform quantum process tomography (QPT).

We now present the experimental implementation of a single-qubit amplitude-damping channel. As the noisy channel is only applied to the probe state, i.e., the polarization degree of freedom of the photons, the longitudinal spatial modes of the photons ($\ket{U}$ and $\ket{D}$) are not affected. The photons on either of the modes encounter the same noisy channel.
In the polarization basis,
the amplitude-damping map is realized by the dual interferometer setup
implemented by splitting the two polarization components and putting independent polarization controls inside a beam displacer (BD) interferometer~\cite{YJO12}.

First a BD whose optical axis is perpendicular to that of the one which is used for preparing hyperentangled states in the state preparation stage splits the two polarization components by directly transmitting the vertically polarized photons and shifting the horizontally polarized photons by a lateral displacement. A half-wave plate (HWP) at $45^\circ$ rotates $\ket{H}$ to $\ket{V}$ and another HWP (H$_\text{A}$) at $\theta_\text{A}$ with $\cos2\theta_\text{A}=-\sqrt{1-\eta}$ applies a rotation $\begin{pmatrix} -\sqrt{1-\eta}&\sqrt{\eta}\\
\sqrt{\eta} &\sqrt{1-\eta}\end{pmatrix}$ on the polarization of photons. The following BD splits and combines the photons due to their polarizations, and the HWPs with certain setting angles are used to rotation the polarization of the photons.

A quartz crystal (QC) with thickness of $28.77$mm~\cite{KGX17} is inserted to reduce the spatial coherence of the photons with different polarizations. 
The sandwich-type HWP-BD-HWP setup works as a 50:50 beamsplitter recombining the photons. Accordingly, with probability $1/2$,
the state emerging from the output port is the desired output state.

\begin{figure}
\includegraphics[width=0.5\textwidth]{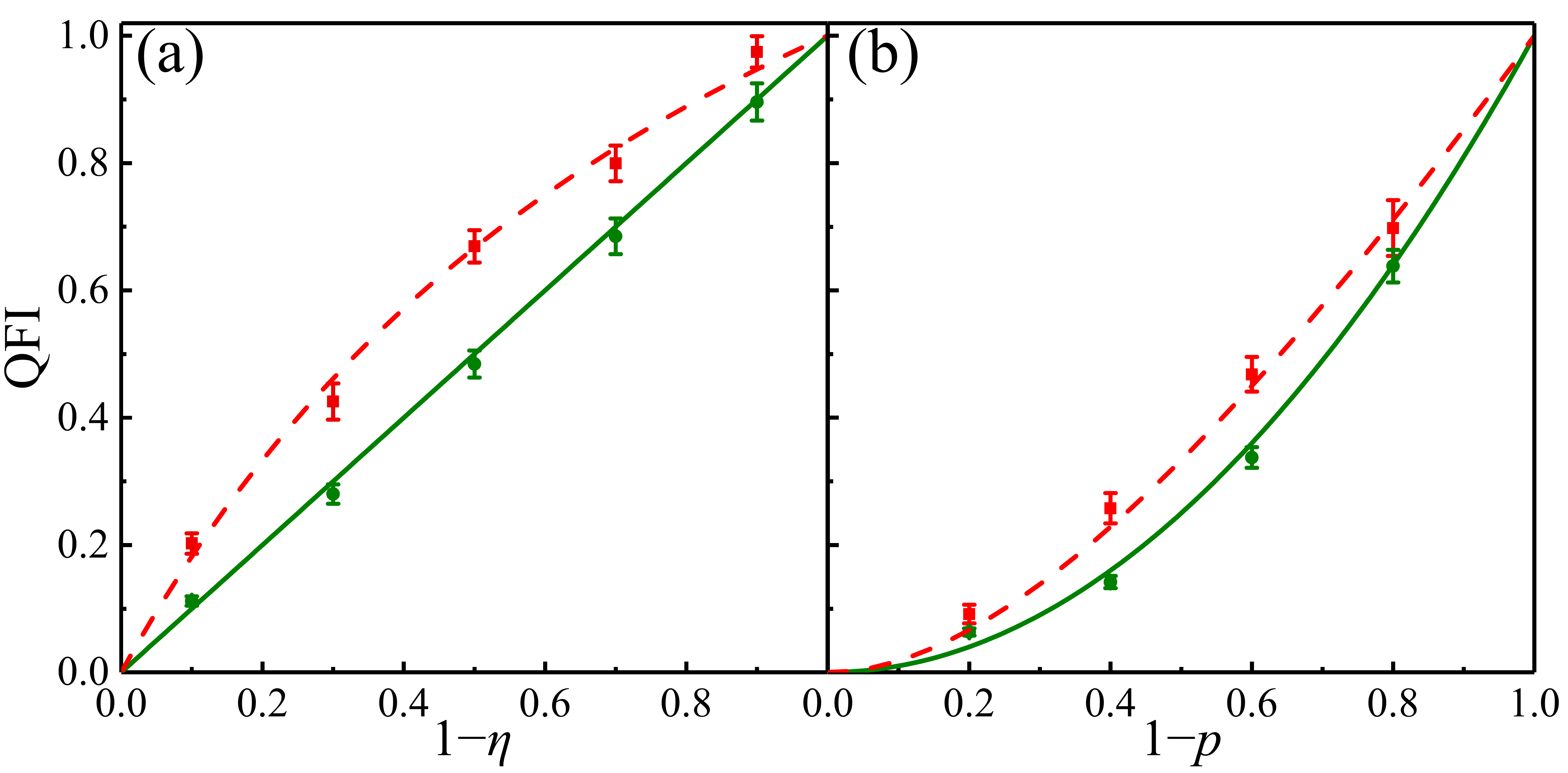}
\caption{Experimental values of QFI. QFI vs~$\eta$ ($p$)
for (a)~amplitude-damping and (b) depolarizing channels. Dashed curves show theoretical predictions of QFI for the entanglement-assisted approach, whereas solid curves are for the optimized single-probe approach.
Data points are experimental results. Error bars are calculated from photon-counting statistics.
}
\label{QFI}
\end{figure}

\begin{figure*}
\includegraphics[width=0.85\textwidth]{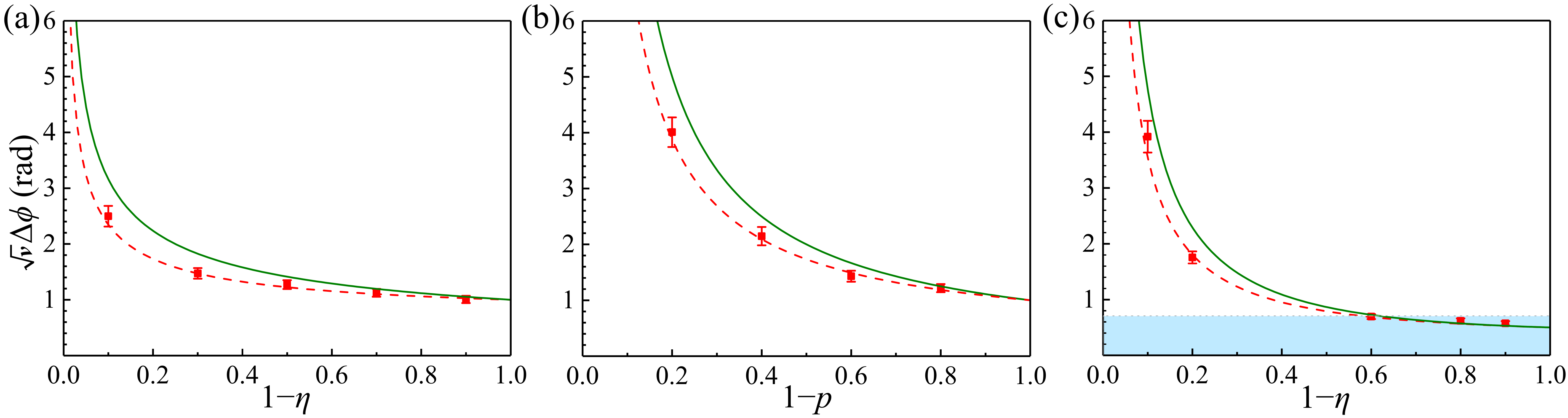}
\caption{Experimental values of the error $\sqrt{\nu}\delta\phi$. The error as a function of the channel noise for single-probe approach in (a) amplitude-damping and (b) depolarizing channels. (c) The result for two-probe approach in the amplitude-damping channel. Dashed curves show theoretical predictions of the error for the entanglement-assisted approach, whereas solid curves are for approaches without assisted entanglement. The grey shadow denotes the shot-noise limit. Data points are experimental results.
Error bars are calculated with the method ``Bootstrap''. Interferometric visibilities of the setups in (a), (b) and (c) are $0.9969\pm0.0006$, $0.9928\pm0.0008$ and $0.9699\pm0.0055$, respectively.
}
\label{error}
\end{figure*}

Furthermore, we can also create a single-qubit space-multiplexed general-Pauli channel~(\ref{eq:PC}) with five BDs and twelve HWPs. Six HWPs (H$_l$ at $\theta_l$, $l=1,\dots,6$) control the ratio of photons in different lateral spatial modes, and three of them at $45^\circ$ (in front of the fifth BD) flip the polarizations and then change the spatial modes of the corresponding photons. Therefore, after the fifth BD, the photons are distributed into four lateral spatial modes according to the parameters $p_i$. For a given desired channel the setting angles $\theta_l$ of the HWPs (H$_l$) are chosen to satisfy the relations
\begin{align*}
\sqrt{p_0}&=\cos2\theta_1\sin2\theta_3=\cos2\theta_2\cos2\theta_4\sin2\theta_6,\\
\sqrt{p_1}&=\sin2\theta_1=-\cos2\theta_2\cos2\theta_4\cos2\theta_6,\\
\sqrt{p_2}&=\cos2\theta_1\cos2\theta_3\cos2\theta_5=\sin2\theta_2,\\
\sqrt{p_3}&=\cos2\theta_1\cos2\theta_3\sin2\theta_5=-\cos2\theta_2\sin2\theta_4.
\end{align*}
Then the last three HWPs at $0^\circ$ and $45^\circ$, respectively, are inserted into different spatial modes and act as Pauli operators $\bm{\Xi}$ on the probe qubit.

Two nonpolarizing beamsplitters (NBSs) recombine the photons in the four lateral spatial modes. To reduce the spatial coherence of the photons, the optical distance~$\varsigma$
between the photons in the different lateral spatial modes should satisfy
$L_\text{coh}<\varsigma<\text{c}\Delta t=0.9$m.
In our experiment,
$\max\varsigma\approx0.6$m.
Hence,
we realize the space-multiplexed general-Pauli channel.

To compare the approaches with and without assisted entanglement,
we realize noisy channels on the probe qubit,
which does not share entanglement with an ancilla.
In our experiment, in both the state preparation and process tomography stages,
the BDs and some WPs are removed from the setup in Fig.~\ref{setup} as no ancillary spatial mode is needed.
In the parametrization stage,
the photons are not distributed into different longitudinal spatial modes.

{\it Experimental results of QFI:-}We present our experimental results for noisy channels and compared the QFI for the single-probe approach with and without assisted entanglement.
Our experimental process matrices $\chi_\text{exp}$ are reconstructed using process fidelity~\cite{XCX08,JAF14}
\begin{equation} F=\frac{\text{Tr}(\chi^\dagger_\text{th}\chi_\text{exp})}{\sqrt{\text{Tr}(\chi^\dagger_\text{exp}\chi_\text{exp})\text{Tr}(\chi^\dagger_\text{th}\chi_\text{th})}}
\label{eq:fidelity}
\end{equation}
to characterize the experimental realization of the noisy channels~\cite{suppl}.
Figure~\ref{channel} shows the experimentally reconstructed $\chi_\text{exp}$ for the amplitude-damping channel with $\eta=0.5$ and the depolarizing channel with $p=0.4$.
Our results exhibit $F\approx1$.
Without assisted entanglement, all the fidelities of the amplitude-damping channel with various parameters are great than $0.9949\pm 0.0007$ and those of the depolarizing channel are greater than $0.9700\pm0.0041$. Whereas with entanglement sharing between the probe and ancilla, all the fidelities of the amplitude-damping channel are greater than $0.9647\pm 0.0003$ and those of the depolarizing channel are greater than $0.9593\pm 0.0016$.

To calculate the QFI, we use the diagonal form of the output state $\rho_\text{exp}(\phi)=\sum_i\lambda_i\ket{\psi_i}\bra{\psi_i}+\rho_\text{noise}$, where $\lambda_i$ and $\ket{\psi_i}$ are the eigenvalues and eigenstates, $\rho_\text{noise}$ is the irrelevant part of the density matrix and is independent of $\phi$~\cite{JMD16}.
With this formula,
we calculate the matrix elements of $A$
in the basis $\{\ket{\psi_i}\}$.

We use the amplitude-damping and depolarizing channels as examples
as usual for decoherence in optical interferometry.
For the amplitude-damping channel, the optimized QFI of the output state is $\frac{\left[2\rho_\text{exp}^{12}(\phi)\right]^2}{\rho_\text{exp}^{11}(\phi)+\rho_\text{exp}^{22}(\phi)}$, and $\frac{\left[2\widetilde{\rho}_\text{exp}^{14}(\phi)\right]^2}{\widetilde{\rho}_\text{exp}^{11}(\phi)+\widetilde{\rho}_\text{exp}^{44}(\phi)}$ for a single-probe input state
and for the entanglement-assisted approach,
respectively,
with $\rho_\text{exp}^{ij}$ a matrix element of $\rho_\text{exp}$.
For the depolarizing channel, without assisted entanglement, the optimized QFI for a single probe is
$\frac{\left[2\rho_\text{exp}^{12}(\phi)\right]^2}{\left[\rho_\text{exp}^{11}(\phi)+\rho_\text{exp}^{22}(\phi)\right]}$. With assisted entanglement, the QFI of the output state of the probe+ancilla system is then $\frac{\left[2\widetilde{\rho}_\text{exp}^{14}(\phi)\right]^2}{\left[\widetilde{\rho}_\text{exp}^{11}(\phi)+\widetilde{\rho}_\text{exp}^{44}(\phi)\right]}+\frac{\left[2\widetilde{\rho}_\text{exp}^{23}(\phi)\right]^2}{\left[\widetilde{\rho}_\text{exp}^{22}(\phi)+\widetilde{\rho}_\text{exp}^{33}(\phi)\right]}$.

As we reconstruct all noisy-channel information via QPT~\cite{NC04,ABJ+03}, the output state for each case is reconstructed. By setting $\phi=0$, we calculate experimental QFI values of the output states. In Fig.~\ref{QFI}, experimental values of the QFI for the amplitude-damping and depolarizing channels either with or without the assisted entanglement are shown.
Our experimental results agree well with theoretical calculations.

Evidently,
for a single probe, in the presence of amplitude-damping noise and depolarizing noise, an entanglement-assisted scheme improves the QFI compared to the unentangled case for all ranges of noise regimes. To illustrate this, we also realize the general-Pauli channel with $p_0=p_2=0.5$ and $p_1=p_3=0$. The experimental value for QFI for the entanglement-assisted approach is $0.984\pm0.045$, which agrees with the theoretical prediction $1$, whereas the optimized QFI for a single probe is $0$.
This represents the case of orthogonal noise when the ancilla approach recovers almost the full information on the phase even in the presence of noise.

{\it Phase estimation:-}For the single-probe approach, the phase $\phi$ to be estimated has been obtained via a unitary map via an additional HWP inserting in the interferometer which causes the optical path difference between photons with different polarizations. The optimal measurement strategy around $\phi\sim 0$ consists in projecting in the polarization-spatial hyperentangled states $(\ket{HU}\pm \text{i}\ket{VD})/\sqrt{2}$. Since no information on $\phi$ is carried on the other bases, for convenience, we choose $\ket{HD}$ and $\ket{VU}$. The projective measurements are realized via a BD, a quarter-wave plate (QWP) at $0$, HWPs at $45^\circ$ and $22.5^\circ$ respectively, and a polarizing beamsplitter (PBS). Coincidences between the outputs and the trigger are detected by single photon avalanche photodiodes (APDs)~\cite{suppl}.

For the amplitude-damping channel, the outcome probabilities of the projective measurements are $P\left[(\ket{HU}\pm \text{i}\ket{VD})/\sqrt{2}\right]=\left[2-\eta\pm2v\sqrt{1-\eta}\sin\phi\right]/4$, $P(\ket{HD})=\eta/2$ and $P(\ket{VU})=0$, where $v$ is the visibility of the interferometer. The optimal measurement is identified by optimising the highest QFI $2v^2(1-\eta)/(2-\eta)$, which proves that the measurement achieves the quantum Cram\'{e}r-Rao bound for the input state. Whereas, for depolarizing channel, the outcome probabilities are $P\left[(\ket{HU}\pm \text{i}\ket{VD})/\sqrt{2}\right]=\left[2-p\pm2v(1-p)\sin\phi\right]/4$, $P(\ket{HD})=p/4$ and $P(\ket{VU})=p/4$ and the corresponding QFI is $2v^2(1-p)^2/(2-p)$, which is always above the single-probe
QFI.

For the two-probe approach, we use the amplitude damping channel as an example. The input state is prepared in a two-photon N00N state $(\ket{HH}+\ket{VV})/\sqrt{2}$. Each probe is affected by an individual amplitude damping channel with the noise parameter $\eta$. With ancillary degree of freedom---spatial modes of two photons, the entanglement-assisted state becomes $(\ket{HUHU}+\ket{VDVD})/\sqrt{2}$. The optimal measurement strategy around $\phi\sim 0$ consists in projecting in the polarization-spatial hyperentangled states $(\ket{HUHU}\pm\text{i}\ket{VDVD})/\sqrt{2}$. No information on $\phi$ is carried on the other $14$ bases. The outcome probabilities of the projective measurements are $P\left[(\ket{HUHU}\pm\text{i}\ket{VDVD})/\sqrt{2}\right]=\left[2-2\eta+\eta^2\mp2v(1-\eta)\sin2\phi\right]/4$, $P(\ket{HDHD}=\eta^2/2)$, $P(\ket{HDVD})=\eta(1-\eta)/2$, $P(\ket{VDHD})=\eta(1-\eta)/2$, and zero for the other projective measurements. The optimal measurement is identified by optimising the highest QFI $8v^2(1-\eta)^2/\left[1+(1-\eta)^2\right]$, which is always above the two-probe approach without assisted entanglement $4v^2(1-\eta)^2/\left[1-\eta+\eta^2\right]$.

To realize the entanglement-assisted single-probe approach, for each of the various noise parameters, data is accumulated for collecting time of $10$s, corresponding to a coincidence count rate of about $20,000$ events per acquisition. Whereas for the entanglement-assisted two-probe approach, the coincidence count rate is about $2,000$ events per acquisition. Totally $100$ values of the phase $\phi$ are collected. The standard deviation of the sample $\delta\phi$ is expected to converge to the ultimate limit established by the quantum Cram\'{e}r-Rao bound in the limit of a large number of repetitions. We use the standard deviation of the sample multiplied by $\sqrt{\nu}$ (here, $\nu$ is the average number of the events) to indicate the error $\sqrt{\nu}\delta\phi$.

Figure~\ref{error} shows the experimental results of the error $\sqrt{\nu}\delta\phi$ as a function of the noise parameters for different approaches in different noisy channels. For the single-probe approach, due to experimental imperfections such as imperfect interferometric visibility of the setup, it is difficult to observe the advantages of the entanglement-assisted approach at low noise. With the noise parameter increasing, the advantages are more obvious. For the two-probe case, the approach of a two-qubit N00N state beats the shot-noise limit both in the noiseless case and at low noise level. The advantage over the classical metrology is affected by noise. Assisted entanglement protects against the noise, especially at high noise level.

{\it Discussion:-}We experimentally realized entangled-assisted quantum metrology
and demonstrated its efficacy through the QFI
for single-qubit amplitude-damping, depolarizing and general-Pauli noisy channels.
Compared to the approach without assisted entanglement,
we observe an enhancement over the noisy cases.
Our achievement relies on replacing time-sharing noisy channels
by space-multiplexed noisy channels using a practical, linear-optical interferometric network.
Our demonstration serves as a foundation for future experimental simulations employing networks of multi-qubit channel simulations.
We use polarization-spatial hyperentangled states encoded in photons,
which are easier to create and control.
Our new approach to entanglement-assisted quantum metrology via a simple linear-optical interferometric network with easy-to-prepare photonic inputs provides a path towards practical quantum metrology.

{\it Note:-}After completing this work,
we learned of related work by the group of Marco Barbieri~\cite{SGM+17}.

{\bf Acknowledgments} We thank Lorenzo Maccone for helpful discussions and appreciate elucidating correspondence with Carlton M. Caves regarding how and why assisted entanglement is an advantage. We acknowledge support by NSFC (Nos.~11474049, 11674056 and GG2340000241), NSFJS (No. BK20160024), the Scientific Research Foundation of the Graduate School of Southeast University and the Open Fund from State Key Laboratory of Precision Spectroscopy of East China Normal University. BCS acknowledges financial support from the 1000-Talent Plan.





\newpage
\begin{widetext}

\renewcommand{\thesection}{\Alph{section}}
\renewcommand{\thefigure}{S\arabic{figure}}
\renewcommand{\thetable}{S\Roman{table}}
\setcounter{figure}{0}
\renewcommand{\theequation}{S\arabic{equation}}
\setcounter{equation}{0}

\section{Supplemental Material for ``Entanglement-enhanced quantum metrology in a noisy environment''}

In this Supplemental Material, we discuss extended-channel quantum Fisher information, optimal probe states under the dynamics with depolarization, intuitive understanding why assisted entanglement helps against noise, as well as some experimental details.

\section{Extended-channel quantum Fisher information}
The action of a quantum channel $\Lambda_\phi=\mathcal{E}\circ\mathcal{U}_\phi$ can always be expressed as its operator-sum representation,
$	\Lambda_\phi\rho
		=\sum_{i}K_i\left(\phi\right)\rho K_{i}^\dagger\left(\phi\right)
$ with Kraus operator $K_i(\phi)$ satisfying $\sum_{i}K_{i}\left(\phi\right)K_{i}^\dagger\left(\phi\right)=\one$. Evidently, this representation is not unique; different sets of linearly independent Kraus operators can be related by unitary transformations~\cite{RJM12}
\begin{equation}
\widetilde{K}_i\left(\phi\right)=\sum_j u_{ij}\left(\phi\right)K_{i}\left(\phi\right),
\end{equation}
where $u_{ij}$ is the element of a unitary matrix $u\left(\phi\right)$ possibly depending on $\phi$.

The single-channel quantum Fisher information is equal to the smallest quantum Fisher information of its purifications $\Lambda_\phi\rho = \text{Tr}_\text{E}\left(\ketbrad{\Psi_{\phi}}\right)$ with $\ket{\Psi_\phi}$ the state of input+environment and the subscript E for tracing out environment~\cite{AH08}
\begin{equation}
	J \left(\Lambda_\phi\rho\right)=\min \limits_{\ket{\Psi_{\phi}}} J\left(\ket{\Psi_{\phi}}\right)
\end{equation}
by minimizing over the state of input+environment $\ket{\Psi_\phi}$.

For a pure input state (not an unreasonable constraint as the optimal input state is always pure~\cite{F17}), different purifications correspond to different Kraus representations of the channel. Moreover, it is enough to parameterize equivalent Kraus representations in Eq.~(S1) with a Hermitian matrix $h$, which is the generator of infinitesimal rotations; i.e., $u\left(\phi\right)=\text{e}^{-\text{i}h\left(\phi-\phi_0\right)}$, in the vicinity of the real value $\phi_0$. This formulation simplifies the optimization problem Eq.~(S2) by revising it as a minimization problem over $h$. Therefore, we obtain the maximal quantum Fisher information after performing the input optimization as~\cite{JR13}
\begin{equation}
	\max \limits_{\rho}J \left(\Lambda_\phi\rho\right)=
4\max \limits_{\rho}\min \limits_{h}\text{Tr}\left(\rho\sum_i\dot{\widetilde{K}}_i^{\dagger}\left(\phi\right)\dot{\widetilde{K}}_i\left(\phi\right)\right)
\end{equation}
with $\dot{\widetilde{K}}_i\left(\phi\right)=\partial_{\phi}\widetilde{K}_i\left(\phi\right)$.

By considering an ancillary system with extended input states involving probe and ancilla, we acquire full information available about $\phi$ imprinted by the map $\Lambda_\phi$ on the extended output state. Then quantum Fisher information of the extended-channel is calculated in a similar way. The map becomes $\widetilde{\rho}\left(\phi\right)=\Lambda_\phi\otimes\one\widetilde{\rho}$, where $\widetilde{\rho}$ denotes the initial pure state of the probe+ancilla system. The quantum Fisher information is
\begin{equation}
	\max \limits_{\widetilde{\rho}}J \left(\Lambda_\phi\otimes \one\widetilde{\rho}\right)=
4\max \limits_{\rho_\text{A}}\min \limits_{h}\text{Tr}\left(\rho_\text{A}\sum_i\dot{\widetilde{K}}_i^{\dagger}\left(\phi\right)\dot{\widetilde{K}}_i\left(\phi\right)\right),
\end{equation}
where $\rho_\text{A}=\text{Tr}_\text{A}\left(\widetilde{\rho}\right)$ is obtained by tracing over the auxiliary space, which leads to the maximization over all mixed states $\rho_\text{A}$. Equation (S4) is exactly Eq.~(S3) with the pure input state replaced by a general mixed one. By maximizing over all mixed states, the extended channel quantum Fisher information can be larger than the unextened one. 
If and only if the optimal $\rho_\text{A}$ is a pure state, assisted entanglement does not help.

\section{Optimal probe states under the dynamics with depolarization}

The depolarizing channel is described by Kraus operators
\begin{equation} K_0=\sqrt{1-\frac{3p}{4}}\Xi_0,K_{1,2,3}=\sqrt{\frac{p}{4}}\Xi_{1,2,3},
\end{equation}
where $\Xi=(\one,X,Y,Z)$ are the Pauli matrices.
Using the method of semi-definite programming~\cite{RJM12}, we find the optimal generator
\begin{align}
 h=\frac{1}{2}\left(\begin{array}{cccc}
                      0 & 0 & 0 & \xi\\
	                  0 & 0 & -\text{i} & 0\\
	                  0 & \text{i} & 0 & 0\\
                      \xi & 0 & 0 & 0
                   \end{array}\right), \xi=\frac{\sqrt{\left(4-3p\right)}}{2-p}.
\end{align}

For the single-probe approach, the optimal input state is $\rho=\ket{+}\bra{+}$, where $\ket{\pm}=(\ket{0}\pm\ket{1})/\sqrt{2}$. Substituting the optimal state and generator into Eq.~(S3), we obtain the maximal quantum Fisher information of the single probe
\begin{equation}
\max \limits_{\rho}J \left(\Lambda_\phi\rho\right)=\left(1-p\right)^2.
\end{equation}

For the entanglement-assisted approach, the optimal reduced state is the maximally mixed state $\rho_\text{A}=\left(\ketbra{0}{0}+\ketbra{1}{1}\right)/2$. The optimal entangled input state in this case is any pure state $\widetilde{\rho}$ with the reduced state equal to $\rho_\text{A}$. The simplest choice of the optimal input state is the maximally entangled state $\widetilde{\rho}=\left(\ket{00}+\ket{11}\right)\left(\bra{00}+\bra{11}\right)/2$~\cite{HMM16}, and the corresponding maximal quantum Fisher information is
\begin{equation}
\max \limits_{\widetilde{\rho}}J \left(\Lambda_\phi\otimes\one\widetilde{\rho}\right)=\frac{2\left(1-p\right)^2}{\left(2-p\right)},
\end{equation} which is always greater than that of the single-probe approach for arbitrary $p\in (0,1)$.

\section{Intuitive understanding why assisted entanglement helps against noise}

The intuitive understanding of how and why the ancilla qubit helps is crucial to making progress on entanglement-assisted metrology. Here,
we provide it for the case of a depolarizing channel.

Figure S1(a) shows the single-probe approach. A Hadamard operator creates the state of the probe qubit $\ket{+}$. With $U_{\phi}=\text{e}^{-\text{i}Z\phi/2}$, the depolarizing channel is
\begin{equation}
\mathcal{E}\odot = \left(1-\frac{3p}{4}\right)\one\odot \one + \frac{p}{4}\left(Z\odot Z+X\odot X+Y\odot Y\right),
\end{equation}
where $\odot$ is a placeholder for the operator which the quantum operation acts on, and the measurement is in the $Y$ basis. Figure S1(b) shows the entanglement-assisted protocol. The Hadamard and controlled-NOT operators together create the entangled state $\ket{\Phi^+}=(\ket{00}+\ket{11})/\sqrt{2}$, and the final measurement is a controlled-NOT followed by $Y\otimes Z$, i.e., $Y$ on the probe qubit and $Z$ on the ancilla qubit.

Then we use the convention that tensor products are written in the order lower-upper. Figure S1(c) shows the second form of the circuit in Fig.~S1(b), in which the first controlled-NOT is moved through the rotation $U_{\phi}$, then moved through the depolarizing channel, combining the second controlled-NOT and then converting the channel to a two-qubit quantum operation
\begin{equation}
\mathcal{F} \odot= \left(1-\frac{3p}{4}\right)\one\otimes \one\odot \one\otimes \one + \frac{p}{4}\left(Z\otimes \one\odot Z\otimes \one+X\otimes X\odot X\otimes X+Y\otimes X\odot Y\otimes X\right).
\end{equation}
The finial measurement is then of $Y\otimes Z$.

The effect of the single-qubit circuit on the state $\ket{+}$ is
\begin{equation}
\mathcal{E}\circ \mathcal{U}_{\phi} \left(\ketbra{+}{+}\right) = \left(1-p\right)U_{\phi}\ketbra{+}{+}U_{\phi}^{\dagger}+\frac{p}{2}\one;
\end{equation}
i.e., the rotation is applied with probability $1-p$, and the qubit is mapped to the maximally mixed state with probability $p$.

The effect of the ancilla-assisted circuit on the state $\ket{+}\ket{0}$ is
\begin{align}
\mathcal{F}\circ \mathcal{U}_{\phi}\otimes \one \left(\ket{+}\ket{0}\bra{0}\bra{+}\right)
&=\left[\left(1-p\right)U_{\phi}\ketbra{+}{+}U_{\phi}^{\dagger}+\frac{p}{4}\one\right]\otimes\ketbra{0}{0}+\frac{p}{4}\one\otimes\ketbra{1}{1}\\
&=\left(1-\frac{p}{2}\right)\left[\left(1-q\right)U_{\phi}\ketbra{+}{+}U_{\phi}^{\dagger}+\frac{q}{2}\one\right]\otimes\ketbra{0}{0}+\frac{p}{4}\one\otimes\ketbra{1}{1},
\end{align}
where
\begin{equation}
q=\frac{p/2}{1-p/2} \qquad \Longleftrightarrow \qquad 1-q=\frac{1-p}{1-p/2}.
\end{equation}

\begin{figure}
\includegraphics[width=0.6\textwidth]{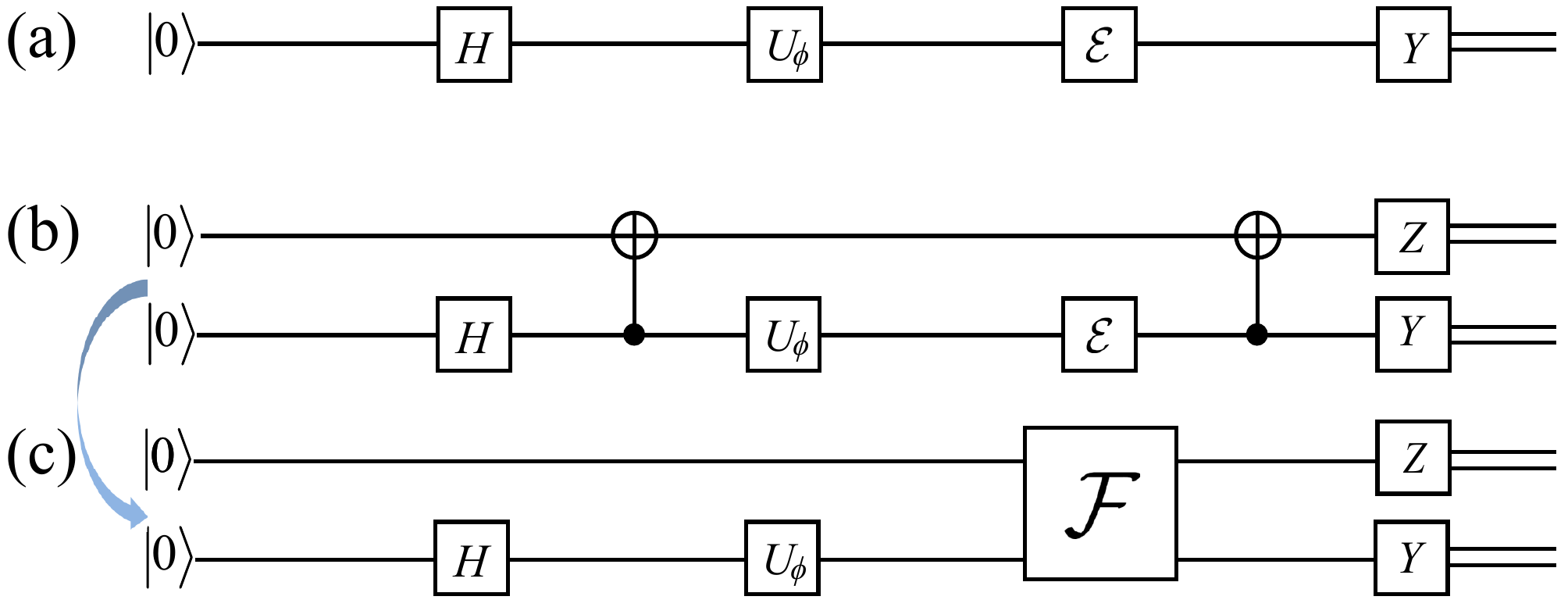}
\caption{(a) Circuit for the single-probe approach. $\mathcal{E}$ is a depolarizing channel. (b) Circuit for the entanglement-assisted protocol. (c) A different form of the circuit for the entanglement-assisted protocol for intuitive understanding why assisted entanglement helps against noise. $\mathcal{F}$ is a specific two-qubit operation.  The lower wire in (b) and (c) is for the probe qubit and the upper wire in (b) and (c) is for the ancilla. The double wire on the right corresponds to a bit from a classical measurement.
}
\end{figure}

Evidently, the form of $\mathcal{F}$ shows that $X$ and $Y$ errors map the main qubit to the maximally mixed state, wiping out the information about $\phi$. This happens just as for the single-qubit circuit, except that a record of when an $X$ or $Y$ error occurs is stored in the ancilla qubit. By monitoring the ancilla qubit, one can discard the random data that results from $X$ or $Y$ errors.

The upshot is that, with probability $1-p/2$, the entanglement-assisted quantum circuit works just like the single-qubit circuit. Compared to the single-qubit circuit, the entanglement-assisted quantum circuit achieves a successful rotation with probability $(1-p/2)(1-q)$, and with probability $(1-p/2)q/2$, maps to the maximally mixed state and with a record stored in the outcome $0$ of the ancilla qubit. As the single-qubit circuit achieves an estimator variance $1/(1-p)^2$, the entanglement-assisted circuit achieves an estimator variance
\begin{equation}
\frac{1}{1-p/2}\frac{1}{(1-q)^2} = \frac{1-p/2}{(1-p)^2},
\end{equation}
which is smaller than $1/(1-p)^2$. That means assisted entanglement helps to achieve an smaller estimator variance compared to the single-probe approach. The term $1-p/2$ in the denominator of the first expression comes from the reduction in the number of trials when one discards the trials that give outcome $1$ on the ancilla qubit.

\section{State preparation}

We prepare single photons in polarization-spatial hyperentangled states for entanglement-assisted single-probe approach~\cite{JTP10,ELL10}. The source consists of a $\beta$-barium-borate (BBO) nonlinear crystal pumped by a CW diode laser, and polarization-degenerate photon pairs at $801.6$nm are generated by a type-I spontaneous parametric down-conversion (SPDC) process. The photon pairs have a coherence length of $L_\text{coh}=214.2\mu$m and spectral bandwidth $\Delta\lambda=3$nm.

Upon detection of a trigger photon, the signal photon is heralded in the measurement setup. This trigger-signal photon pair is registered by a coincidence count at two single-photon APDs with a $\Delta t=3$ns time window. Total coincidence counts are about $20,000$ over a collection time of $10$s. The probe is encoded in the horizontal $\ket{H}$ and vertical $\ket{V}$ polarizations of the heralded single photons.

After passing through a PBS followed by a HWP and a QWP, the single photons are prepared in an arbitrary single-qubit state. The longitudinal spatial modes $\ket{U}$ and $\ket{D}$ represent the basis states of the ancilla. A birefringent calcite BD whose optical axis is cut so that horizontally polarized light is directly transmitted and vertical light undergoes a longitudinal displacement into a neighboring mode, acts as an effective controlled-NOT gate on the polarizations and the spatial modes and prepare the initial state into a polarization-spatial hyperentangled state $\alpha\ket{HU}+\beta\ket{VD}$ ($|\alpha|^2+|\beta|^2=1$ and $\alpha,\beta\neq 0$).

Whereas, for entanglement-assisted two-probe approach, polarization-entangled photon pairs are used to prepare the four-qubit hyperentangled state. Similarly, entangled photons in $(\ket{HH}+\ket{VV})/\sqrt{2}$ are also generated via type-I SPDC. Two $\beta$-BBO crystals and a following titled HWP (H$_\text{C}$) placed right after two joint $\alpha$-BBO crystals are used to compensate the walk-off between photons with horizontal and vertical polarizations. Each photon passes through a BD and then a four-qubit polarization-spatial hyperentangled state $(\ket{HUHU}+\ket{VDVD})/\sqrt{2}$ is generated. Total coincidence counts are about $2,000$ over a collection time of $10$s.

\section{Accuracy of the noisy channel simulation}

To verify accuracy of the noisy channel simulation,
we reconstruct the process matrices of the channels via two-qubit QPT~\cite{NC04,ABJ+03}.
The action of a generic channel operating on a probe qubit is
\begin{equation}
	\mathcal{E}(\widetilde{\rho})=\sum_{n,m,n',m'=0}^3	\chi_{nmn'm'}(\Xi_n\otimes\Xi_m)\widetilde{\rho}(\Xi_{n'}\otimes\Xi_{m'}),
\end{equation}
where~$\chi_{nmn'm'}$ completely characterizes the process.
To determine $\mathcal{E}$ we first choose some fixed states $\{\widetilde{\rho}\}$,
which form a basis for the set of operators acting on the state space of the probe+ancilla system. Each state is then subject to the process $\mathcal{E}\otimes\one$,
and quantum state tomography is used to determine the output state $(\mathcal{E}\otimes\one) \widetilde{\rho}$.

A total of sixteen initial states $\widetilde{\rho}_l$, $l=1,\dots,16$,
and sixteen measurements on a two-qubit state of the probe+ancilla system are needed. These states are generated by PBS, BD and WPs.
The HWP (H$_{\text{S}1}$), 
and QWP (Q$_{\text{S}1}$) are used to control the ratio and relative phase between the photons in the upper and lower modes, respectively, whereas H$_{\text{S}2}$ is used to control the ratio between the photons with different polarizations and Q$_{\text{S}2}$ is for the relative phase.
Measurements are performed in the bases
\begin{align}
	\left\{\ket{H},\ket{V},\frac{\ket{H}-\text{i}\ket{V}}{\sqrt{2}},
	\frac{\ket{H}+\ket{V}}{\sqrt{2}}\right\}\nonumber\\
	\otimes\left\{\ket{U},\ket{D},\frac{\ket{U}-\text{i}\ket{D}}{\sqrt{2}},
	\frac{\ket{U}+\ket{D}}{\sqrt{2}}\right\}.
\end{align}

After reconstructing the process matrices, we use process fidelity in Eq.~(\ref{eq:fidelity}) to characterize the experimental realization of the noisy channels.

\section{Projective measurements for realizing phase estimation}

For entanglement-assisted single-probe approach, the optimal measurement strategy around $\phi\sim0$ is projecting the output state into four basis states: \begin{equation*}
\left\{\frac{1}{\sqrt{2}}(\ket{HU}\pm \text{i}\ket{VD}),\ket{HD},\ket{VU}\right\},
\end{equation*} respectively. The projective measurements can be realized via a BD, a QWP, several HWPs and a PBS.
A sandwich-type setup, i.e., HWP(at $45^\circ$)-BD-HWP(at $45^\circ$) separate the photons in the states $\ket{VU}$ and $\ket{HD}$ into the uppermost and lowest modes, and combine the photons in the states $\ket{HU}$ and $\ket{VD}$ into the middle mode. In the middle mode, a QWH at $0^\circ$ following by a HWP at $22.5^\circ$ applies a rotation on the polarization states, i.e.,
\begin{equation*}
                \frac{1}{\sqrt{2}}\left(\ket{H}-\text{i}\ket{V}\right)\longrightarrow\ket{H}, \text{        } \frac{1}{\sqrt{2}}\left(\ket{H}+\text{i}\ket{V}\right)\longrightarrow\ket{V}.
\end{equation*}
Finally the PBS projects the photons in the middle mode into two basis states $(\ket{HU}\pm \text{i}\ket{VD})/\sqrt{2}$. Coincidences between the
outputs and the trigger are detected by APDs. The outcome
probabilities of projecting the state in the basis $\left\{(\ket{HU}\pm \text{i}\ket{VD})/\sqrt{2},\ket{HD},\ket{VU}\right\}$ depend on the coincidences between two of APDs (D$_0$, D$_\text{R}$), (D$_0$, D$_\text{L}$), (D$_0$, D$_\text{H}$) and (D$_0$, D$_\text{V}$), respectively.

For entanglement-assisted two-probe approach, the optimal measurement strategy around $\phi\sim0$ is projecting the output state into sixteen basis states:
\begin{align*}
\Big\{&\frac{1}{\sqrt{2}}\big(\ket{HUHU}\pm\text{i}\ket{VDVD}\big),\ket{HUHD},\ket{HUVU},\ket{HUVD},\ket{HDHU},\ket{HDHD},\ket{HDVU},\\
&\ket{HDVD},\ket{VUHU},\ket{VUHD},\ket{VUVU},\ket{VUVD},\ket{VDHU},\ket{VDHD},\ket{VDVU}\Big\},
\end{align*}
respectively. Similar to the entanglement-assisted single-probe approach, the projective measurements here are realized via BDs, WPs, NBSs, and a PBS. We use a multi-channel coincidence counting system that records all possible combinations of two-photon detection
events occurring coincidentally across $12$ APDs (D$_1$,\dots,D$_{12}$). The outcome
probabilities of projecting the state in the bases depends on the combinations of coincidences between pair of APDs (D$_1$,\dots,D$_{12}$). The corresponding relation is shown in Table S1.

\begin{table}[tb]
\small
\centering
\caption{The corresponding relations between the projective measurements and the combinations of coincidences between pair of APDs. Here, `/' denotes `or', and `,' between `()' means `and'. For example, $\left(\text{D}_5 / \text{D}_6,\text{D}_9 / \text{D}_{10}\right),\left(\text{D}_7 / \text{D}_8,\text{D}_{11} / \text{D}_{12}\right)$ means that the outcome
probability of projecting the state in the bases $\left(\ket{HUHU}+\text{i}\ket{VDVD}\right)/\sqrt{2}$ depends on the coincidences between pairs of APDs such as (D$_5$,D$_9$), (D$_5$,D$_{10}$), (D$_6$,D$_9$), (D$_6$,D$_{10}$), (D$_7$,D$_{11}$), (D$_7$,D$_{12}$), (D$_8$,D$_{11}$), and (D$_8$,D$_{12}$). The superscript `$*$' denotes that the probability of projective measurement depends on the doubled coincidences. That is because in some case two photons happen to be in the same port of the NBS with half of the probability, which can not be recorded in the experiment. Thus we need to double the coincidences for the rest cases to represent the correct outcome probability of projective measurement.}
\begin{tabular}{c||c|c|c|c}
  \hline\hline
  Basis state & $\left(\ket{HUHU}+\text{i}\ket{VDVD}\right)/\sqrt{2}$ & $\ket{HUHD}$ & $\ket{HUVU}$ & $\ket{HUVD}$ \\
  \hline
  Coincidences & $\left(\text{D}_5 / \text{D}_6,\text{D}_9 / \text{D}_{10}\right),\left(\text{D}_7 / \text{D}_8,\text{D}_{11} / \text{D}_{12}\right)$ & $\left(\text{D}_9 / \text{D}_{10} / \text{D}_{11} / \text{D}_{12},\text{D}_4\right)$ & $\left(\text{D}_9 / \text{D}_{10} / \text{D}_{11} / \text{D}_{12},\text{D}_3\right)$ & $(\text{D}_9,\text{D}_{10})^*,(\text{D}_{11},\text{D}_{12})^*$ \\
  \hline\hline
  Basis state & $\ket{HDHU}$ & $\ket{HDHD}$ & $\ket{HDVU}$ & $\ket{HDVD}$ \\
  \hline
  Coincidences & $\left(\text{D}_2,\text{D}_5 / \text{D}_6 / \text{D}_7 / \text{D}_8\right)$ & $\left(\text{D}_2,\text{D}_4\right)$ & $\left(\text{D}_2,\text{D}_3\right)$ & $\left(\text{D}_2,\text{D}_9 / \text{D}_{10} / \text{D}_{11} / \text{D}_{12}\right)$ \\
  \hline\hline
  Basis state & $\ket{VUHU}$ & $\ket{VUHD}$ & $\ket{VUVU}$ & $\ket{VUVD}$ \\
  \hline
  Coincidences & $\left(\text{D}_1,\text{D}_5 / \text{D}_6 / \text{D}_7 / \text{D}_8\right)$ & $\left(\text{D}_1,\text{D}_4\right)$ & $\left(\text{D}_1,\text{D}_3\right)$ & $\left(\text{D}_1,\text{D}_9 / \text{D}_{10} / \text{D}_{11} / \text{D}_{12}\right)$ \\
  \hline\hline
  Basis state & $\left(\ket{HUHU}-\text{i}\ket{VDVD}\right)/\sqrt{2}$ & $\ket{VDHD}$ & $\ket{VUVU}$ & $\ket{VDHU}$ \\
  \hline
  Coincidences & $\left(\text{D}_5 / \text{D}_6,\text{D}_{11} / \text{D}_{12}\right),\left(\text{D}_7 / \text{D}_8,\text{D}_9 / \text{D}_{10}\right)$ & $\left(\text{D}_5 / \text{D}_6 / \text{D}_7 / \text{D}_8,\text{D}_4\right)$ & $\left(\text{D}_5 / \text{D}_6 / \text{D}_7 / \text{D}_8,\text{D}_3\right)$ & $(\text{D}_5,\text{D}_6)^*,(\text{D}_7,\text{D}_8)^*$  \\
  \hline\hline
\end{tabular}
\end{table}

\end{widetext}
\end{document}